\documentclass{iaus}
\usepackage{graphics}

\title[Be/Oe stars and LGRBs] 
{Massive Oe/Be stars at low metallicity: Candidate progenitors of long
GRBs?}

\author[C.\ Martayan, D.\ Baade, J.\ Zorec, et al.]   
{C.\ Martayan$^{1,2}$, 
D.\ Baade$^3$,
J.\ Zorec$^4$
Y.\ Fr\'emat$^5$,
J.\ Fabregat$^6$
\\
\and S.\ Ekstr\"om$^7$}

\affiliation{$^1$ESO Chile; 
$^2$GEPI-Observatoire de Meudon, France;
$^3$ESO Germany;
$^4$Instritut d'Astrophysique de Paris, France; 
$^5$Royal Observatory of Belgium, Brussels, Belgium;
\\
$^6$Valencia University, Spain;
$^7$Geneva Observatory, Switzerland
}

\pubyear{2011}
\volume{272}  
\pagerange{1--2}
\jname{Active OB stars: structure, evolution, mass loss and critical limits}
\editors{C.\ Neiner, G.\ Wade, G.\ Meynet \& G.\ Peters, eds.}
\begin{document}
\maketitle

\begin{abstract}
At low metallicity the B-type stars rotate faster than at higher metallicity, typically in the SMC.
As a consequence, it was expected a larger number of fast rotators in the SMC than in the Galaxy, 
in particular more Be/Oe stars. With the ESO-WFI in its slitless mode, the SMC open clusters were 
examined and an occurence of Be stars 3 to 5 times larger than in the Galaxy was found.
The evolution of the angular rotational velocity seems to be the main key on the understanding 
of the specific behaviour and of the stellar evolution of such stars at different metallicities.
With the results of this WFI study and using observational clues on the SMC WR stars and massive stars, 
as well as  the theoretical indications of long gamma-ray burst progenitors, we identify the 
low metallicity massive Be and Oe stars as potential LGRB progenitors.
Therefore the expected rates and numbers of LGRB are calculated and compared to the observed ones, 
leading to a good probability that low metallicity Be/Oe stars are actually LGRB progenitors.

\end{abstract}

\firstsection 
\section{Long GRBs and the collapsar model}
In the collapsar model for Ib/c supernovae (\cite[Woosley
1993]{woosley}), matter with high specific angular momentum is
accreted by the already-formed black hole with a short delay.  The
intermediate disk structure gives rise to the formation of
relativistic jets emitting the so-called long gamma-ray bursts
(GRBs). The gamma-ray radiation is strongly relativistically beamed
with opening angles of order 10 degrees.

In order for a rapidly rotating core to be present, quasi-homogeneous
chemical evolution is essential.  Stellar evolution models show that
in stars with very high initial rotation rates mixing is faster than
chemical gradients are built up by nuclear burning (for more details
see \cite[Yoon \& Langer 2005]{YoonLanger}).

The angular momentum still available at this late phase of stellar
evolution is the larger, the less mass and, therefore, angular
momentum the progenitor star has lost during its lifetime.  Since
radiatively driven mass loss from massive stars decreases with
metallicity, this has led to the notion that low-metallicity
star-forming regions should be preferred hosts of GRBs.  In fact,
observations are beginning to support this picture (e.g., \cite[Modjaz
et al.\ 2008]{Modjaz}).  Moreover, the GRB rate per unit mass
increases with redshift $z$.  But it is not clear whether this
requires any explanation other than the general increase in star
formation with redshift (up to z$\sim$2).

\section{Oe/Be stars}
In the Milky Way, the near-main sequence stars with the highest
rotation velocities are Oe/Be stars.  Moreover, many of them rotate at
$\sim$90\% of the critical rate.  Two arguments predict a larger
fraction of nearly critically rotating massive stars in
low-metallicity environments: (i) The mass (and angular momentum) loss
rates are lower and (ii), all else being equal, the radii are smaller.
Various comparisons of open clusters in the Magellanic Clouds and the
Galaxy have concluded (cf.\ \cite[Martayan, Baade \& Fabregat
2010]{Martayan}) that the abundance of Oe/Be stars increases with
decreasing metallicity $Z$.  If one keeps in mind that rapid rotation
only is a necessary, but not also sufficient, condition for a rapidly
rotating O/B star to become an Oe/Be star, this can pass as a
confirmation of the expectation.  In any event, because of their
emission lines, Oe/Be stars probably are more reliable tracers of a
population of rapidly rotating OB stars than broad photospheric lines
would be.  Therefore, Oe/Be stars more massive than
$\sim$18\,$M_{\odot}$ (metallicity dependent; \cite [Yoon, Langer \&
Norman 2006]{Yoon}) qualify as candidates, also at low $Z$.  Note that,
perhaps, Oe/Be stars do not quite reach the theoretical upper mass
limit for the progenitors of GRBs (\cite[Yoon, Langer \&
Norman 2006]{Yoon}).

\section{GRBs and Oe/Be stars in a synopsis}
The strong relativistic beaming makes GRBs observable out to extreme
redshifts.  But it also implies that much fewer nearby GRBs are
detected.  Therefore, while it is still relatively straightforward to
determine the number of Oe/Be stars per SMC mass and also the space
density of such environments as a function of redshift can be
estimated, statistics with just a few dozen entries (GRBs) pose
serious problems.  Nevertheless, using a distribution of beam angles
determined from observations, 3-6 long GRBs are predicted at $z \leq
0.2$ per 11-year period while the observed number is 8 (cf.\
\cite[Martayan et al.\ 2010]{Martayanetal}).  Since the agreement is better 
than 2 $\sigma$, there is no need to fine tune adjustment parameters
such as the range in spectral type, from which GRB progenitors are
drawn, or the number, type, and star-formation level of the assumed
host galaxies.  But the excellent agreement could still be
coincidental.

\section{Discussion}
In addition to the challenges already mentioned, a number of other
questions still warrant further examination:
\begin{list}{$\bullet$}{\itemsep=0mm\labelsep=2mm\topsep=0mm}
\item
Do Oe/Be stars really preserve their large intial angular momentum
throughout their evolution?  After all, they regularly eject
rotationally supported disks.  This behavior is not observed in On/Bn
stars, i.e.\ rapidly rotating O/B stars without emission lines.
\item
Are there other differences in the evolution of Be and Bn stars? 
\item
Is the SMC a good proxy for GRB-forming environments?  
\item
How incomplete are the observations, especially of GRBs?
\item
Can candidate progenitors other than Oe/Be stars reach the GRB state
along different evolutionary paths?
\end{list}

\end{document}